\documentclass[traditabstract]{aa}
\usepackage{amssymb}
\usepackage{natbib}
\usepackage{graphicx}
\bibpunct{(}{)}{;}{a}{}{,}

\begin{document}

\title{Spectral line profiles changed by dust
scattering in heavily obscured young stellar objects}

\author
{V. P. Grinin\inst{1,2,3} \and L.V.\,Tambovtseva\inst{1,3} \and
 G.\,Weigelt\inst{3}}

\institute {Pulkovo Astronomical Observatory of the Russian
Academy of Sciences, Pulkovskoe shosse 65, St. Petersburg, 196140,
Russia\\
e-mail: grinin@gao.spb.ru, lvtamb@mail.ru
 \and  The V.V.\,Sobolev Astronomical Institute of the St.
Petersburg University, Petrodvorets, St. Petersburg, 198904,
Russia \and Max-Planck-Institut f\"{u}r
Radioastronomie, Auf dem Hugel 69, D-53121 Bonn, Germany\\
e-mail:weigelt@mpifr-bonn.mpg.de}

\offprints{V. P. Grinin}

\date{Received\, 6 June 2012/\,Accepted 2 July 2012}

\titlerunning{ }

\authorrunning{Grinin et al.}

\abstract{It is known that scattering of radiation by
circumstellar dust can strongly change the line profiles in
stellar spectra. This hampers the analysis of spectral lines
originating in the emitting regions of heavily obscured young
stars. To calculate the line profile of the scattered radiation,
we suggest to use the approximation of remote scattering
particles. This approximation assumes that the scattering dust
grains are at a distance from the star that is much larger than
the characteristic size of the emitting region. Using this method,
we calculated the line profiles of several simple models. They
show the H$\alpha$ line profiles of Herbig AeBe stars in the
presence and absence of motionless or moving dust.}

\maketitle

\section{Introduction}
Some amount of radiation scattered by circumstellar (CS) dust is
present in the radiation of many young stars and is the main
source of their linear polarization (Bastien \& Landtsreet, 1979).
In most cases, the contribution of this radiation is very small
and does not exceed a few percent. Exceptions are three groups of
young stellar objects whose scattered radiation can substantially
exceed the direct stellar radiation. First, these are young
stellar objects (YSO) embedded in opaque gas and dust envelopes.
According to the current classification, these are objects of
Class I (White et al. 2007). The extinction toward some of them
exceeds 20$^m$ (see, e.g., Beck et al. 1991). Young stars
surrounded by circumstellar disks seen edge-on belong to the
second group. In these stars, the direct radiation is strongly
attenuated due to the extinction in the disk, and one can observe
them only via scattered radiation (Padgett et al. 1999). The T
Tauri star HH 30 is a prototype of these objects (Burrows et al.
1996). The third group consists of the UX Ori type stars. The
photometric activity of these young stars is caused by the
nonhomogeneous structure of their circumstellar disks and their
small inclination to the line of sight (Grinin et al. 1991; Natta
et al. 1999). Opaque fragments of the CS disk intersect the line
of sight from time to time. During these events, the direct
radiation of the star decreases while the contribution of the
scattered radiation increases, which is testified by the growth of
the linear polarization of the star (Grinin et al. 1991).

\begin{figure}[h]
\begin{center}
\includegraphics[width=7cm,angle=-90]{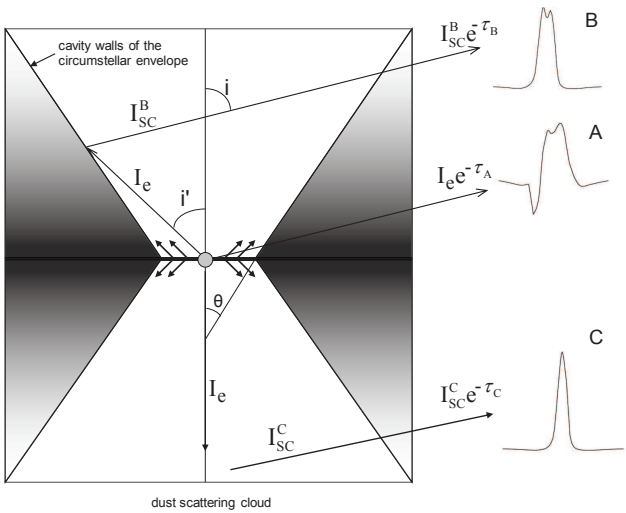}
\caption{Sketch of a deeply embedded young star with a rotating
accreting disk, disk wind, and an envelope (cocoon) with
low-density polar cavities. Dust particles are assumed to be near
the polar axis and walls of the cavity. Possible profiles of
emission lines are shown for several viewing conditions: direct
(A) and scattered (B and C) radiation from the central object,
which consists of the star, and its surrounding emitting region
(e.g., disk and disk wind region). Here $I_{e}$ is the emergent
radiation from this central object, $I_{sc}^B$ and $I_{sc}^C$ is
the radiation scattered by the dust particles in cases B and C
respectively, $\tau_A$, $\tau_B$ and $\tau_C$ are the optical
thickness of the circumstellar envelope at different heights above
the disk ($\tau_B \ll \tau_A$ and $\tau_C \ll \tau_A$), $i$ is the
viewing angle, $i'$ is the angle between the polar axis and
direction to the scattering particle, and $\theta$ is a
geometrical parameter.}\label{ske1}
\end{center}
\end{figure}

In all these cases, an observer obtains the information on the
stellar spectrum by observing the radiation scattered by the
circumstellar dust. It is well known that in the coordinate system
of the dust particle, the photon frequency before and after
scattering is the same. However, the resulting scattered-light
spectrum is not the same as the spectrum in the absence of
scattering.

As noted by Appenzeller, Bertout \& Stahl (2005), scattered light
is not a precise source of information about obscured objects.
These authors observed several T Tauri stars with edge-on disks.
The analysis of their spectra showed that the disks are completely
opaque at visible wavelengths and that light from the central
objects reaches an observer only via scattering layers above and
below the disk planes. According to these authors, light could be
scattered either by matter located along the rotation axis or in
the disk atmosphere. As a result, the spectrum of scattered
radiation can differ from the spectrum of the CS emitting region.
Grinin, Mitskevich \& Tambovtseva (2006) considered the broadening
of photospheric lines due to the scattering by the CS dust at the
puffed-up inner rim of the accretion disks. This effect can be
observed in UX Ori type stars during eclipses by opaque fragments
of CS disks. The transformation of the spectral line profiles due
to dust scattering in expanding dusty envelopes can also be
observed in the spectra of evolved stars (see Lefevre 1992, and
references therein).

Below we consider several model situations typical for embedded
YSOs (Fig.~\ref{ske1}) and show how the profiles of spectral lines
broadened by motion in the emitting gas (e.g., disk wind) can
change after scattering by circumstellar dust particles.
\section{Scattering by the motionless dust}
The effect of scattering by dust particles depends on the geometry
and kinematics of the line-formation region and the scattering
region (see Fig.~\ref{ske1}). If the spectral line arises, for
example, in a flattened rotating envelope or accretion disk, the
line profiles of the direct and scattered radiation can be very
different: observer A, for example, would observe a wide
double-peaked or P Cygni line profile from the accretion disk
and/or disk wind. However, if the direct view to the emitting
region is blocked, then observer B would observe a narrow single
or double-peaked line. If there is scattering dust near the polar
axis of the disk far away from the disk, then observer C would
observe a very narrow line, since the radial motion seen at the
position of the scattering dust is almost zero. These three simple
examples show that the observed spectrum depends on both the
geometry and kinematics of the emitting region itself and the
distribution of the scattering particles in the neighborhood of
the star.

\begin{figure}[b]
\begin{center}
\includegraphics[width=5cm, angle=-90]{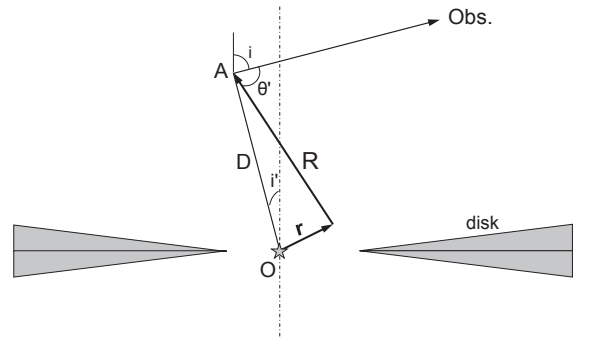}
\caption{Geometric parameters of the considered model (a central
cross-section of the CS region). }\label{ske2}
\end{center}
\end{figure}

In the general case when the spectral line originates in a moving
gas with an arbitrary velocity field, the expression for the
intensity $I_{sc}(\nu)$ at the frequency $\nu$ of the radiation
scattered at point $A$ (Fig.~\ref{ske2}) by a single motionless
(in the coordinate system of the star) particle is
\begin{eqnarray}\label{issc}
      I_{sc}(\nu) =
      \pi a^2 Q_{sc}\int\limits_{\textbf{V}}S(\textsf{r})
      \phi(\textsf{r},\nu-\nu_0
      \frac{\textsf{v}_R(\textsf{r})}{c})
   \times\\ \nonumber
      e^{-\tau(\nu,\textsf{r})-\tau_d({\textsf{r})}}
\kappa(\textsf{r})f(\theta')\frac{d\textbf{V}}{R^2}
\end{eqnarray}

Here $\textbf{V}$ is the volume occupied by the emitting gas, $a$
is the radius of the particle, $Q_{sc}$, $f$, and $\theta'$ are
the efficiency factor for scattering, phase function, and the
scattering angle, respectively, $S(\textsf{r})$ is the source
function for the transition considered at the point $\textsf{r}$
of the emitting region, $\phi$ is the profile of the absorption
coefficient normalized to the unity, $\textsf{v}_R(\textsf{r})$ is
the projection of the velocity of the emitting gas in the point
$\textsf{r}$ onto the vector $\textbf{R}$, which connects this
point with the scattering particle at point $A$,
\begin{equation}
\tau_d = \tau'_d + \tau''_d,
\end{equation}
where $\tau'_d$ is the optical thickness between point
$\textsf{r}$ and point $A$ caused by the CS extinction, $\tau''_d$
is the optical thickness between point $A$ and the observer;
$\kappa$ is the integrated line opacity in the considered spectral
line, and $\tau(\nu,\textsf{r})$ is the line optical depth at
point $\textsf{r}$ and for the frequency $\nu$ in the direction of
point $A$
\begin{equation}\label{tau}
\tau(\nu,\textsf{r}) =
\int_{0}^{R}\kappa(\textsf{r}')\phi(\nu-\nu_0\frac
{\textsf{v}_R(\textsf{r}')}{c})dR'\,.
\end{equation}
Here the integration is performed along the vector $\textbf{R}$
linking the scattering particles with the point $\textsf{r}$;
$\textsf{r'}$ and $R'$ are the integration variables. We assume a
complete redistribution in the line frequencies in the reference
frame of the atom and use the Doppler profile $\phi$ in the
calculations of $\tau$ and the line intensity.

In order to take into account scattering by all particles, we have
to integrate the expression (1) over the total volume occupied by
the dust, as well as to consider all sizes and types of dust
particles participating in scattering. The solution of this
problem faces substantial difficulties. Therefore, below we
consider the simplified model when scattering grains are at some
fixed distance $D = |\textbf{D}|$ from the star, and $D$ is much
larger than the characteristic size of the emitting region. This
approximation is quite realistic in many cases since the region of
the permitted line formation is usually much more compact than the
region occupied by the CS dust.

\subsection{Approximation of remote scattering particles}
In this case, we can use the following simplifications: i) in
integral (1), it is possible to replace $R^{-2}$ by $D^{-2}$ and
thus to extract $D^{-2}$ from the integral; ii) in the phase
function $f(\theta')$, it is possible to replace $\theta'$ by
$\psi$, where $\psi$ is the angle between vector $\textbf{D}$ and
direction from the point $A$ to the observer, and also extract it
from the integral. We can do the same with the optical thickness
$\tau_d$ if we use the optical thickness between the points $O$
(location of the star) and $A$ instead of $\tau'_d$. After these
simplifications we obtain the asymptotic equation for single
scattering by the elementary volume at point $A$, which can be
written in the following form:
\begin{equation}\label{3}
      I^A_{sc}(\nu) = \frac{\pi a^2} {D^2}\,Nf(\psi)\,Q_{sc}
      \,I_\infty(\nu)\,e^{-\tau_d},
\end{equation}
where N is the number density of particles, $I_\infty(\nu)$ is the
over the whole emitting region integrated intensity at frequency
$\nu$ in the direction to the remote observer whose line of sight
coincides with the vector \textbf{D}:
\begin{equation}\label{4}
I_\infty(\nu) = \int\limits_{\textbf{V}}S(\textsf{r})
      \phi(\textsf{r},\nu-\nu_0\frac{\textsf{v}_R(\textsf{r})}{c})
e^{-\tau(\nu,\textsf{r})}\kappa(\textsf{r})d\textbf{V}.
\end{equation}
Since the scattering particle is assumed to be motionless in the
star coordinate system, the frequency of the scattered radiation
in any direction will coincide with the frequency of the incident
radiation, and since the values of $f(\psi)$ and $Q_{sc}$ before
the integral are practically constant within the profile of the
spectral line, the value $I_\infty(\nu)$ will determine the line
profile in the spectrum of the scattered radiation. It is
important that this profile is the same for all scattering angles
and depends only on the angle $i'$.

It should be stressed that the intensity of the radiation in the
expression given above corresponds to the case of single
scattering. Since we assume here that the dust particles are
motionless in the star coordinate system, the photon frequency
will be conserved in the next scattering. Thus, the first
scattering plays a key role in the transformation of the line
profile in the considered case. It should be noted, that in the
optical region of the spectrum, albedo of scattering by CS dust is
about 0.5 (see, e.g., Natta \& Whitney 2000). Therefore, the first
scattering gives the main contribution to the intensity of the
scattered radiation.

\subsection{The H$\alpha$ line in the spectrum of the scattered
radiation} As an example, let us consider the following model: the
emission spectrum of the young star is formed in the axially
symmetric disk wind, which starts from the surface of the
accretion disk. The star and the emission region are surrounded by
an optically thick axially symmetric envelope with cavity (see
Fig. 1), whose polar region is transparent for the radiation. The
spectrum of the radiation scattered by the dust on the surface of
the cavity walls and by dust in the polar region is observed as
the spectrum of the star.

For the calculations of $I^A_{sc}(\nu)$ of the H$\alpha$ line, we
used the model of the magnetocentrifugal disk wind of Ae/Be Herbig
stars reported by Grinin \& Tambovtseva (2011). A description of
this model is given in the cited paper and the Appendix.

\begin{figure}
\begin{centering}
\includegraphics[width=90mm,angle =-90]{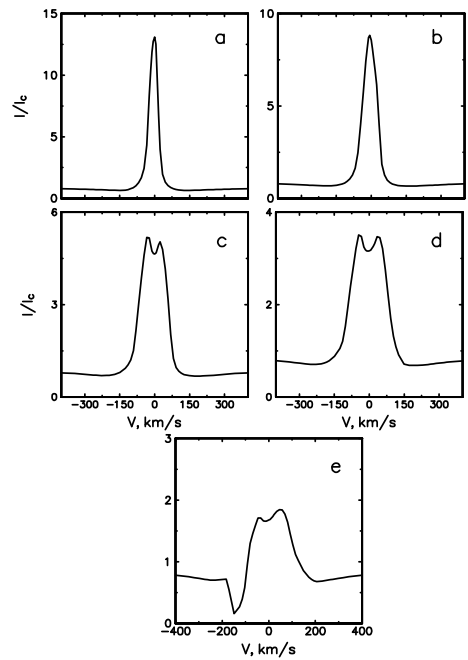}
\caption{The profile of the H$\alpha$ line in the spectrum of the
radiation scattered at point A (a,b,c,d) and the spectrum of
direct radiation (e). The parameters of the wind model are
described in the text and the Appendix. The angle $i'$  between
the symmetry axis of the disk and the vector $\textbf{D}$ are
5$^\circ$ (a), 20$^\circ$ (b) ,40$^\circ$(c), and 60$^\circ$ (d).
The angle between the line of sight and the symmetry axis of the
disk is $i = 80^\circ$ (e).}\label{mod4}
\end{centering}
\end{figure}
The profiles of the H$\alpha$ line in the spectrum of the
scattered radiation are calculated for several cases where the
angle $i'$ between the symmetry axis of the wind and the direction
to the scattering particle ranged from 5 to 60 degrees
(Fig.~\ref{mod4}a-d), and the angle between the line of sight and
the symmetry axis of the disk was $i = 80^\circ$. Figure
~\ref{mod4}e shows the H$\alpha$ line profile which the observer
could see in the absence of scattering. We see that this profile
(broadened by the motion of the emitting gas) has a $P$ Cygni
shape (instead of pure emission) and it is wider than the profiles
observed after scattering at the point A. This example shows that
even in the case of the motionless dust, the profile of the
spectral line of the direct radiation and the radiation after
scattering can strongly differ.
\section{Scattering by the moving dust}
\begin{figure}[h]
\begin{center}
\includegraphics[width=6cm]{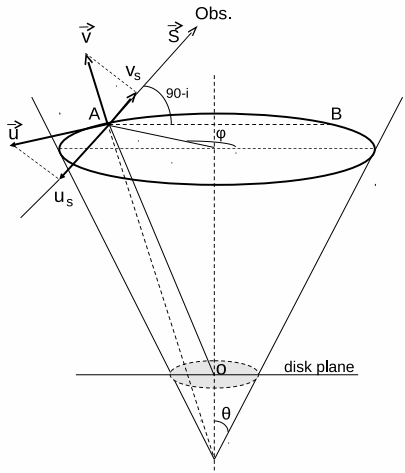}
\caption{The sketch of a dust particle at point $A$ moving with
the velocity components $\textbf{v}$ and $\textbf{u}$ and
scattering the photons in the direction to an
observer.}\label{proj}
\end{center}
\end{figure}

In the reality, the dust particles are moving. The motion may be
purely Keplerian rotation of the CS disk or purely radial
(bi-conical) outflow, or their combination, like in the dusty disk
wind (Safier 1993; K\"{o}nigl \&  Salmeron, 2011). Scattering of
photons by moving dust provides an additional Doppler shift of the
photon frequency, which leads to an additional deformation of the
line profiles. In this case, in the equation~(\ref{3}) we have to
replace $\nu$ by $\nu'$, where
\begin{equation}\label{5}
\nu' = \nu + [v_s+u_s-v\cos{(\theta-i')}]/c \,
\end{equation}
Here $v_s$ and $u_s$ are the projections of the radial ($v$) and
rotation ($u$) velocity components at point $A$ on the line of
sight \textbf{s}:
\begin{equation}\label{6}
v_s = v(\cos{\theta}\cos{i} + \sin{\theta}\sin{i}\cos{\varphi})\,,
\end{equation}
\begin{equation}\label{7}
u_s = - u\sin{i}\sin{\varphi},\,
\end{equation}
where $\varphi$ is the azimuthal coordinate angle of point $A$
(see Fig.~\ref{proj}).

The term $-v\cos{(\theta-i')}/c$ in Eq.~(\ref{5}) takes into
account the Doppler shift of the frequency $\nu$ due to the motion
of the scattering particles relatively to the emitting region.

The intensity of the scattered radiation at the frequency $\nu$ in
the direction to an observer is the integral over the whole volume
V:
\begin{equation} I_{sc}(\nu) =\int_V I^A_{sc}(\nu')\,d\textbf{V}\,
\end{equation}
Below we consider two limiting cases: a) radial motion of the dust
particles with constant velocity $v$ and tangential velocity $u =
0$, and b) tangential motion of the dust with the velocity $u$ and
radial velocity $v = 0$. The intensity of the scattered radiation
from all elementary segments of the scattering volume with fixed
coordinates $\theta$, $D$ is:

\begin{equation}
I_{sc}(\nu) \propto \int_0^{2\pi} I^A_{sc}(\nu')\,d\varphi,
\end{equation}
where $I^A_{sc}(\nu)$ is determined from equation~(\ref{3}) but
with changes described above (Eqs.(\ref{5}) - (\ref{7})).

When calculating the line profile of the scattered radiation on
the base of this formula, we assumed for simplicity that $\theta =
i'$, the optical thickness  between the scattering particles and
the observer $\tau'' \ll 1$, and the scattering is isotropic.

\begin{figure*}
\begin{centering}
\includegraphics[width=5cm,angle =-90]{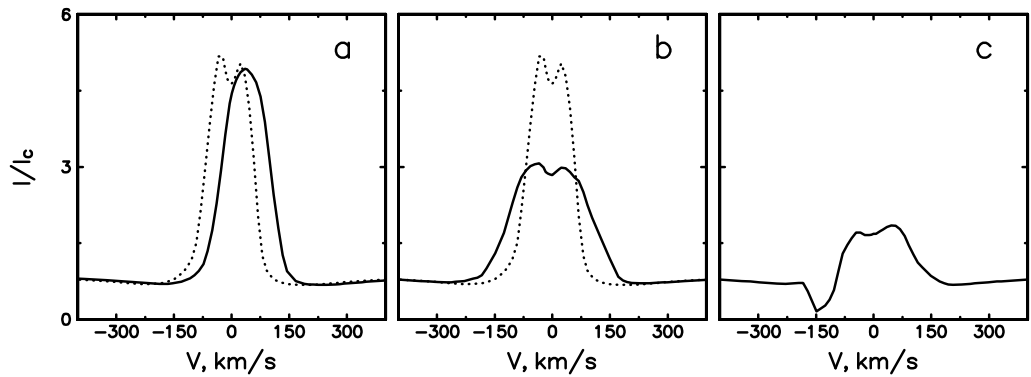}
\caption{The profile of the H$\alpha$ line in the spectrum of the
scattered radiation with (solid) and without (dots) assuming dust
motion: a) line profile transformation due to a radial outflow
with velocity $v$ = 100 km\,s$^{-1}$; b) line profile
transformation due to rotation ($u$ = 100 km\,s$^{-1}$); c)line
profile of direct radiation in the absence of dust scattering. The
angle between the symmetry axis and the direction to the
scattering particle is $i' = 40^\circ$, the angle between the
symmetry axis of the disk and the line of sight is $i =
80^\circ$.}\label{d1}
\end{centering}
\end{figure*}

\begin{figure}
\begin{centering}
\includegraphics[width=5cm,angle =-90]{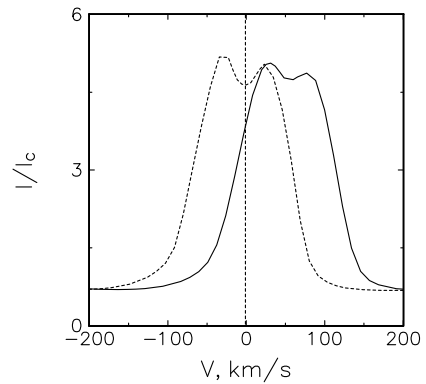}
\caption{The profile of the H$\alpha$ line in the spectrum of the
scattered radiation with (solid) and without (dots) radial outflow
of the dust with $v$ = 100 km\,s$^{-1}$. The contribution of the
back wall is only considered here. Therefore, the angle $\varphi$
in Eq.~(\ref{7}) is varied from $\pi/2$ to $3\pi/2$. The other
parameters are the same as in Figure~\ref{d1}. See text for more
details.}\label{d2}
\end{centering}
\end{figure}

The results of the calculations for the limiting cases described
above are presented in Figs.~\ref{d1} and~\ref{d2}. According to
Fig.~\ref{d1}, a radial motion of the dust from the star provides
a red shift of the line profile. This shift will be larger if the
main contribution to the scattered light is provided by the back
wall of the conical cavity (Fig.~\ref{d2}), which is observed in
some heavily obscured stars (see, e.g., Preibisch et al. 2003).
The rotation does not shift the line profile, but smooths it and
makes it broader (Fig.\ref{d1}b).

These examples show that the effect of motion of the CS dust leads
to more divers transformations of the line profiles compared to
the case of motionless dust. In the both cases considered above,
the line profile of the scattered radiation strongly deviates from
the profile of the direct radiation in the absence of scattering
presented in Fig.\ref{d1}c.

\section{Summary and concluding remarks}
The presented calculations show that the profiles of spectral
lines arising in moving CS gas can change when scattering by CS
dust takes place. The distortion can appear even if the CS dust is
not moving in the coordinate system of the star. The lines have
more diverse forms in the case of the moving dust. This is
important to keep in mind when analyzing emission spectra of
heavily obscured young stars, and it is especially important for
the comparison of spectral lines at different wavelength (e.g.,
H$\alpha$ and Br$\gamma$), since the dust extinction depends on
the wavelength. In heavily obscured young objects, the H$\alpha$
line is observed via the scattering radiation in practically all
cases. The contribution of the scattered light to the Br$\gamma$
line is less important if $A_V \leq 7-8^m$. However, in young
objects with $A_V \gg 8^m$, even the Br$\gamma$ is predominatingly
observed in the scattered light and its profile can also be
distorted by scattering. In these cases, only observations of mid-
and far-infrared spectral lines may avoid these scattering
effects.

The changes of the spectral line profiles due to the scattering by
the CS dust is also valid for the photospheric spectrum of the
star itself since its radiation propagates in a similar same way
to an observer as the radiation of the CS emitting region. As a
result, the measured radial and rotational velocities of the
heavily obscured star can differ from their real values.

It should be noted, that in the case when the spectral line is
formed in a spherically symmetric stellar wind, the spectrum of
radiation scattered by a single particle at an arbitrary remote
point will be the same for any scattering angle and will coincide
with the spectrum of the direct radiation of the star. This is the
only kind of model where the profiles of the spectral lines do not
change in the case of dust scattering.

Finally, we emphasize that highly obscured YSOs are, of course,
not the only objects in which the scattered radiation can be an
important component of the observed radiation. Similar conditions
exist in active galaxy nuclei observed under certain inclinations
(see Antonucci \& Miller 1985, Antonucci 1993; Cohen et al. 1999;
Smith et al. 2005, and references there). In these objects, the
scattered radiation arises due to the Thomson scattering on free
electrons and also due to scattering by dust particles in the
dusty torus and the Narrow Line region. Spectropolarimetric
observations have played an important role in studies of the
physical properties of these objects. Similar spectropolarimetric
observations would also be very useful in studies of heavily
obscured YSOs.

\begin{acknowledgements} This research has been supported in part
by the program of the Presidium of RAS N 21 and grant N.\,Sh.-
1625.2012.2.
\\
\end{acknowledgements}

\clearpage

{\Large \bf Appendix: the disk wind model}

\begin{figure}[b]
\begin{centering}
\vspace{10mm}
\includegraphics[width=55mm, angle =-90]{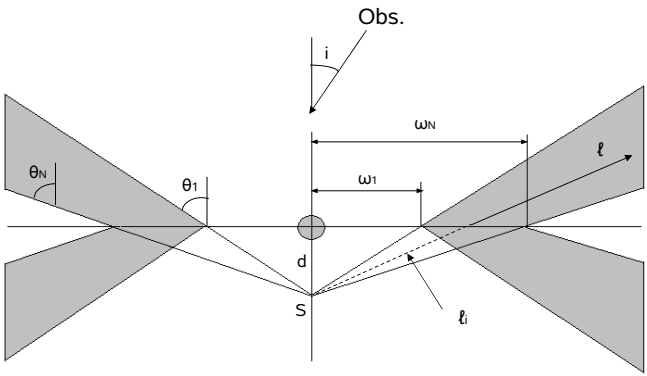}
\vspace*{1mm} \caption{Sketch of the geometry of the disk-wind
model adopted.}\label{ske3}
\end{centering}
\end{figure}

\vspace{5mm}
For the description of the disk wind, we use a
coordinate system ($l$, $\theta$, $\omega$) centered at point S
(see the sketch in Fig.~\ref{ske3}). In MHD models of the disk
wind, the inclination angle of the first streamline with respect
to the disk plane is typically assumed to be around $60^\circ$
(Blandford \& Payne 1982). However, in the case of the hot Herbig
Be stars, the strong stellar radiation pressure can bend down the
streamlines and make the disk wind flatter.
The kinematical model described below is typical for disk-wind
models adopted for different astrophysical objects with accretion
disks including young stars stars (see e.g. Kurosava et al. 2006),
and references there). It is assumed that the tangential velocity
of the wind changes along streamlines as:
\begin{equation}
u(\omega) = u_K(\omega_i)\,(\omega/\omega_i)^{-1}
\end{equation}
where $\omega$ = $l\, \sin{\theta}$ is the distance of the point
($l,\theta$) from the rotation axis and $\omega_i$ =
$l_i\,\sin{\theta}$; $u_K\,(\omega_i$) = $(G\,M_*/\omega_i)^{1/2}$
at the point ($\omega_i$) at the base of streamline $i$ .

The radial velocity $v$ increases along the streamlines as
\begin{equation}
v(l) = v_0 + (v_{\infty} - v_0)\,(1 - l_{i}/l)^{\beta}
\end{equation}
where $v_0$ and $v_{\infty}$ are the initial and terminal values
of the radial velocity and $\beta$ is a parameter. We  adopt for
simplicity $v_\infty$ = $f\,u_K(\omega_i)$, where $u_K(\omega_i)$
is the Keplerian velocity at distance $\omega_i$ from the disk
axis, and $f$ is a parameter. As in the Kurosawa et al. (2006)
paper we used $f$ = 2.

The local mass-loss rate per unit area of the disk, $\dot m_w$, is
a function of the cylindric radius $\omega$. For the description
of $\dot m_{w}$, we use a simple power law
\begin{equation}
\dot m_{w}(\omega)\sim \omega^{-\gamma},
\end{equation}
where $\gamma$ is the input parameter. The other input parameter
is the total mass loss rate integrated over all the disk wind
ejection region (from both sides of the disk) $\dot M_w$.

Using the continuity equation, we calculated the distribution of
the number density in each point of the disk wind. The
calculations of the ionization state and the number densities of
the atomic levels were performed for the isothermal disk wind
model with the account of both collision and radiative processes
on the base of the Sobolev approximation (see Grinin \&
Tambovtseva (2011) for more details).

The radiation of the disk wind was calculated for the following
model parameters: the mass loss rate $\dot M_w = 10^{-9} M_\odot$
yr$^{-1}$; the electron temperature $T_e$ = 8000 K, $M_* = 2.5
M_\odot$, $R_* = 2.4 R_\odot$, $\gamma$ = 1, the disk wind
launching region is located between 0.1 and 1 AU. The half opening
angle of the disk wind $\theta_1 = 60^\circ$ (see
Fig.~\ref{ske3}). The radiation of the star is described by the
Kurucz model with the effective temperature $T_{ef} = 10^4$ K and
$\log g = 4$.
\end{document}